\documentclass[twocolumn,superscriptaddress,showpacs,prl]{revtex4}

\usepackage{graphicx}
\usepackage{bm}

\begin{document}

\title{Quantum critical behavior near a density-wave instability in an
  isotropic Fermi liquid}

\author{Andrey V. Chubukov} \affiliation{Department of Physics, University
  of Wisconsin, Madison, WI 53706} 
  
  \affiliation{Condensed Matter Theory Center,
  Department of Physics, University of Maryland, College Park, MD
  20742-4111}

\author{Victor M. Galitski} \affiliation{Condensed Matter Theory Center,
  Department of Physics, University of Maryland, College Park, MD
  20742-4111}

\affiliation{Kavli Institute for Theoretical Physics, University of
  California, Santa Barbara, CA 93106}

\author{Victor M. Yakovenko} \affiliation{Condensed Matter Theory Center,
  Department of Physics, University of Maryland, College Park, MD
  20742-4111}

\date{cond-mat/0405344, v.2 8 February 2005, Phys.~Rev.~Lett.\ {\bf
    94}, 046404 (2005)}


\begin{abstract}
  We study the quantum critical behavior in an isotropic Fermi liquid
  in the vicinity of a zero-temperature density-wave transition at a
  finite wave vector $q_c$.  We show that, near the transition, the
  Landau damping of the soft bosonic mode yields a crossover in the
  fermionic self-energy from $\Sigma(k,\omega)\approx\Sigma(k)$ to
  $\Sigma(k,\omega)\approx\Sigma(\omega)$, where $k$ and $\omega$ are
  momentum and frequency.  Because of this self-generated locality,
  the fermionic effective mass diverges right at the quantum critical
  point, not before, i.e., the Fermi liquid survives up to the
  critical point.
\end{abstract}


\maketitle

\paragraph{Introduction.}  

An isotropic Fermi liquid may experience various intrinsic quantum
instabilities.  They are characterized by divergence of the
corresponding static susceptibility and emergence of a gapless bosonic
mode in the collective, two-particle excitation spectrum.  The
instability point at zero temperature, occurring at a particular value
of a control parameter such as electron concentration, is called the
quantum critical point (QCP).  Near QCP, the interaction between the
soft bosonic mode and low-energy fermions often leads to singular
behavior of the fermionic self-energy $\Sigma(k,\omega)$ and
divergence of the fermionic effective mass $m^*$.  A well-known
example is the divergence of $m^*$ near ferromagnetic instability
\cite{Doniach}, which occurs at the wave vector $q=0$.

In this paper, we study the divergence of $m^*$ in an isotropic Fermi
liquid near a zero-temperature charge- or spin-density-wave
instability occurring at a nonzero wave vector $q_c\leq2k_F$, where
$k_F$ is the Fermi momentum. We argue that, for $q_c\neq0$, the
behavior of $m^*$ near QCP is rather tricky, and the analysis requires
extra care.

We consider a model in which fermions $\psi_k$ interact by exchanging
a soft bosonic mode $V(q)$
\begin{equation}
\label{Hint}
  {\cal H}_{\rm int} = - \sum_{\bm{q},\bm{k},\bm{k}'}
  \psi_{\bm{k}+\bm{q}}^{\dagger} \psi_{\bm{k}'-\bm{q}}^{\dagger} 
  V (\bm{q})\, \psi_{\bm{k}'} \psi_{\bm{k}}.
\end{equation}
The soft bosonic mode is peaked at the wave vector $q_c$ and can be
in either spin or charge channel
\begin{equation}
\label{chi_st}
  V(q)\approx\frac{g}{\xi^{-2}+(|\bm q|-q_c)^2}.
\end{equation}
Here $g$ is an effective interaction constant, and $\xi$ is the
correlation length, which diverges at QCP as a function of electron
concentration or pressure.

Interaction between soft bosonic modes (\ref{chi_st}) was studied by
Brazovskii \cite{Brazovskii} in the context of a crystallization
transition in an isotropic liquid.  Dyugaev \cite{Dyugaev} applied the
model (\ref{Hint}) and (\ref{chi_st}) to explain enhancement of the
effective mass and specific heat in liquid $^3$He, arguing that $^3$He
is close to a spin-density-wave transition. Ref.\ \cite{Schmalian}
utilized the Brazovskii model to describe a magnetic transition in
MnSi, where a finite $q_c$ is likely caused by the
Dzyaloshinskii-Moriya interaction. The model (\ref{Hint}) and
(\ref{chi_st}) also applies to itinerant electrons in the vicinity of
a ferromagnetic instability -- a small but finite $q_c$ appears there
as a result of an effective long-range interaction between fermions
due to the $2k_F$ Kohn anomaly \cite{Belitz-Chitov-Maslov}.  Refs.\
\cite{Shaginyan,Yakovenko,Galitski} proposed that the model
(\ref{Hint}) and (\ref{chi_st}) can explain the enhancement and
possible divergence of the effective mass observed experimentally in
the two-dimensional electron gas (2DEG) \cite{Kravchenko}, as well as
in $^3$He films \cite{Casey}.  In the scenario of Refs.\
\cite{Shaginyan,Yakovenko,Galitski}, the instability at $q_c$ develops
as a precursor to the Wigner crystal in 2DEG or to the crystallization
transition in $^3$He films.

The effective mass $m^*$ is extracted from the fermionic self-energy
$\Sigma(k,\omega)$ defined by the Dyson equation $G^{-1}(k,\omega)=
i\omega-\varepsilon_k-\Sigma(k,\omega)$, where $G(k,\omega)$ is the
fermion Green's function, and $\varepsilon_k=v_F(k-k_F)$ is the bare
fermion dispersion counted from the chemical potential.  The
derivatives of $\Sigma(k,\omega)$ determine the renormalization factor
$Z^{-1}=1+i\partial_{\omega}\Sigma$ and the effective mass
$m^*/k_F=1/v_F^*=Z^{-1}(v_F+\partial_k\Sigma)^{-1}$, where $v_F$ and
$v_F^*$ are the bare and renormalized Fermi velocities.  Divergence of
$m^*$ can be caused either by $i\partial_{\omega}\Sigma\to\infty$
(and, hence, $Z\to0$), or by $\partial_k\Sigma\to-v_F$.  In the former
case, $m^*$ would diverge at QCP, but not earlier, while in the latter
case, the divergence of $m^*$ would generally occur at a finite
distance from QCP. There exist other scenarios
\cite{Zverev-DasSarma-Krotscheck} for the divergence of $m^*$, which
do not invoke a density-wave transition, but we will not discuss them
here.

The interplay between $\partial_{\omega}\Sigma$ and $\partial_k\Sigma$
depends on whether $\Sigma(k,\omega)$ predominantly depends on
momentum $k$ or on frequency $\omega$.  The two alternative scenarios
for the model of Eqs.\ (\ref{Hint}) and (\ref{chi_st}) where $\Sigma$
depends only on $k$ or only on $\omega$ were advocated in Refs.\
\cite{Shaginyan,Yakovenko,Khodel} and Refs.\ \cite{Dyugaev,Schmalian},
correspondingly.  In this paper, we show that the behavior of $\Sigma$
in an isotropic Fermi liquid near a density-wave transition is
actually rather involved. At some distance from QCP, $\Sigma(k,\omega)
\approx\Sigma (k)$.  However, the frequency dependence of $V$,
generated by the Landau damping, makes $\Sigma(k,\omega)$
predominantly $\omega$-dependent in the immediate vicinity of QCP.  We
show that, when the fermion-boson interaction $g$ is smaller than the
Fermi energy $E_F\sim v_Fk_F$, the crossover from
$\Sigma(k,\omega)\approx\Sigma(k)$ to
$\Sigma(k,\omega)\approx\Sigma(\omega)$ is separated from the
weak-to-strong coupling crossover near QCP.  The latter occurs when
the dimensionless coupling $\lambda\sim(g/E_F)(\xi k_F)\propto\xi$
becomes of the order of 1.  On the other hand, the crossover from
$\Sigma(k)$ to $\Sigma(\omega)$ occurs at a small
$\lambda\sim(g/E_F)^{1/2}\ll1$, where $\Sigma$ is still small, and
$\partial_k\Sigma$ does not reach $-v_F$.  Once $\Sigma (k, \omega)$
becomes $\Sigma (\omega)$, only $\partial_\omega \Sigma$ matters,
i.e., $m^* = (Zv_F)^{-1}$ diverges with $Z^{-1}$ at QCP, but not
earlier.  We present calculations in the 2D case, but the results are
qualitatively valid also in the 3D case.

\paragraph{Momentum-dependent self-energy $\Sigma(k)$ away from QCP.}  

\begin{figure}
\includegraphics[width=0.48\columnwidth]{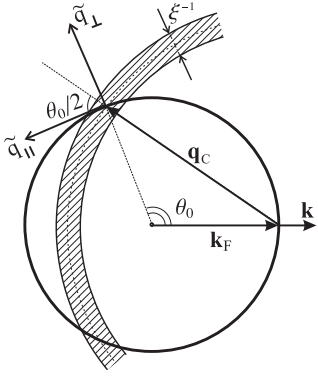}
\caption{The solid circular line represents the Fermi surface.  The
  ring of radius $q_c$ and width $\xi^{-1}$ represents the effective
  interaction (\ref{chi_st}) via a soft bosonic mode.  A fermion with
  the momentum $\bm k$ close to $\bm k_F$ strongly interacts with the
  two ``hot spots'' obtained by intersection of the Fermi circle and
  the interaction ring.  The vector components $\tilde q_\perp$ and
  $\tilde q_\|$ are perpendicular and parallel to the Fermi surface at
  the hot spots.}
\label{fig:circle}
\end{figure}

In the Hartree-Fock approximation, the exchange diagram with the
effective interaction (\ref{chi_st}) gives
\begin{eqnarray}
  \Sigma(\bm k,\omega) &=& \int{d\Omega\,d^2q\over(2\pi)^3}\,
  G(\bm{k}+\bm{q},\Omega+\omega)\, V(\bm{q})
\label{S1}
\nonumber \\
  &=& \int{d^2q\over(2\pi)^2}\,n_F(\varepsilon_{\bm{k}+\bm{q}})
  \, V(\bm{q}),
\label{S2}
\end{eqnarray}
where $n_F(\varepsilon)$ is the Fermi distribution function.  The
integration over $\bm{q}$ in Eq.\ (\ref{S2}) is restricted by the
conditions that the vector $\bm{k}+\bm{q}$ lies inside the Fermi
circle and the vector $\bm{q}$ belongs to the ring of radius $q_c$ and
width $\xi^{-1}$ centered at the vector $\bm{k}$, as shown in Fig.\
\ref{fig:circle}.  Clearly, $\Sigma$ in Eq.\ (\ref{S2}) does not
depend on $\omega$, but it does depend of $k$, because the area of the
ring inside the Fermi circle changes with $k$.

The derivative of Eq.\ (\ref{S2}) with respect to $\bm k$, taken at
$\bm k=\bm k_F$, is given by the integral along the Fermi circle
\begin{equation}
  \frac{\partial\Sigma}{\partial\bm k}
  = - \int{d^2q\over(2\pi)^2}\,
  \delta(\varepsilon_{\bm{k}+\bm{q}})\, 
  \bm v_{\bm{k}+\bm{q}}\,V(\bm{q}).
\label{dS/dk}
\end{equation}
where $\delta(\varepsilon)$ is the Dirac delta-function, and $\bm
v_{\bm{k}+\bm{q}}$ is the Fermi velocity at $\bm{k}+\bm{q}$.  For
large $\xi$, the integral (\ref{dS/dk}) comes from the vicinity of the
two ``hot spots'' $\bm q_c$ obtained by intersection of the Fermi
circle and the circle of radius $q_c$ centered at the point $\bm k_F$
on the Fermi circle (see Fig.\ \ref{fig:circle}).  Decomposing the
deviation from the hot spot $\tilde{\bm q}=\bm q-\bm q_c$ into
$(\tilde q_\perp,\tilde q_\|)$, as shown in Fig.\ \ref{fig:circle},
and integrating over $\tilde q_\|$ first, we obtain $\partial_{\bm
k}\Sigma=\hat{\bm k}\,\partial_k\Sigma$, where
\begin{equation}
  \frac{\partial\Sigma}{\partial k}=-\lambda\,v_F\cos\theta_0, \quad
  {m^* \over m} = {v_F \over v_F^*} = {1 \over 1-\lambda\cos\theta_0}.
\label{pert}
\end{equation}
Here $\theta_0$ is the angle between $\bm k_F$ and $\bm k_F+\bm q_c$
in Fig.\ \ref{fig:circle}, such that $\sin(\theta_0/2)=q_c/(2k_F)$,
and
\begin{equation}
  \frac{\lambda}{2} = \frac{1}{v_F}\int \frac{d\tilde
  q_\|}{(2\pi)^2}\, V (\bm q)= \frac{g\,\xi}{4\pi
  v_F\cos(\theta_0/2)}.
\label{lambda}
\end{equation}
We see that, if $\lambda\cos\theta_0>0$ (which implies
$q_c<\sqrt{2}p_F$ for $\lambda>0$), the effective mass increases with
$\lambda$ and nominally diverges at $\lambda\cos\theta_0=1$, while
$\xi$ is still finite, and QCP is not reached yet.

\paragraph{Crossover to the frequency-dependent self-energy 
  $\Sigma(\omega)$.}

To verify whether Eq.\ (\ref{pert}) holds up to $\lambda\sim1$, where
$m^*$ diverges, we need to go beyond the Hartree-Fock approximation
and include the full fermionic and bosonic propagators and vertex
corrections into the self-energy diagram. We assume and then verify
that, for $g/E_F\ll 1$, the higher-order corrections predominantly
renormalize $V$ in Eq.\ (\ref{S1}), while vertex corrections and the
renormalization of $G$ can be neglected for arbitrary $\lambda$.  The
renormalization of $V (q)$ originates from the electron polarizability
\begin{equation}
\label{P}
  \Pi(q, \Omega) = \int \frac{d^2k\,d\Omega}{(2\pi)^3} \, {1 \over
  i(\omega+\Omega)-\varepsilon_{\bm{k}+\bm{q}}} \, {1 \over i\omega
  -\varepsilon_{\bm{k}}}
\end{equation}
via the relation $V^{-1}(q,\Omega)=V^{-1}(q)+\Pi(q,\Omega)$.  For $q$
near $q_c$, the static part of $\Pi(q,\Omega)$ comes from the fermions
with high energies and we assume that it is already included into Eq.\
(\ref{chi_st}), which implies that $\xi$ is the exact (renormalized)
correlation length.  The dynamical part of $\Pi(q_c,\Omega)$ comes
from low energies and describes the Landau damping of the bosonic mode
due to its decay into particle-hole pairs.  For $\Omega\ll v_Fq_c$,
Im$\Pi(q_c,\Omega)\propto|\Omega|/v_F^2$.  Inserting the Landau
damping term into Eq.\ (\ref{chi_st}), we find
\begin{equation}
\label{chi}
  V(q,\Omega)\approx{g \over \xi^{-2}+(q-q_c)^2+\gamma|\Omega|},
  \quad \gamma\sim{g \over v_F^2}.
\end{equation}
Re-evaluating $\Sigma(k,\omega)$ in Eq.\ (\ref{S2}) for the full
$V(q,\Omega)$, we find that it now depends on both $k$ and $\omega$.
Notice that causal analytical properties require that the interaction
constant $g$ in Eq.\ (\ref{chi}) must be positive: $g>0$ \cite{g<0}.

We present the results for $\Sigma (k,\omega)$ first and discuss the
details of calculations later.  For small $\epsilon_k = v_F (k-k_F)$
and $\omega$, we then obtain
\begin{eqnarray}
  \Sigma(k,0) &=& -\lambda\cos\theta_0\,\varepsilon_k\,h_k(\eta), 
\label{SIG-k} \\
  \Sigma(k_F,\omega) &=& -\lambda\, i\omega\,h_\omega(\eta),
\label{SIG2} \\
   \eta &=& \gamma E_F \xi^2 \sim \lambda^2\,(E_F/g).
\label{eta}
\end{eqnarray}
Here and below we subtracted the renormalization of the chemical
potential from $\Sigma(k,\omega)$, i.e., redefined
$\Sigma(k,\omega)\equiv\Sigma(k,\omega)-\Sigma(k_F,0)$.  The functions
$h_k(\eta)$ and $h_\omega(\eta)$ have the following asymptotic
behavior.  For $\eta\ll1$, $h_k(\eta)=1+O(\eta)$ and
$h_\omega(\eta)\sim\eta\ln(1/\eta)$, i.e.,\ the momentum-dependent
piece in $\Sigma$ almost coincides with Eq.\ (\ref{pert}), while the
frequency dependence of $\Sigma$ is weak.  This is natural, because
small $\eta$ corresponds to small bosonic damping $\gamma$.  However,
for $\eta\gg1$, we find the opposite behavior:
$h_k(\eta)\propto\eta^{-1/2}\ll1$ and
$h_\omega(\eta)=1+O(\eta^{-1/2})$.  In this case, the momentum
dependence of $\Sigma$ is weak compared with the Hartree-Fock
approximation, while its frequency dependence is strong.  Moreover,
$\Sigma$ does {\it not} depend on $\gamma$ explicitly.  Thus, the
limiting forms of $\Sigma(k,\omega)$ are
\begin{eqnarray}
  \Sigma(k,\omega) &\approx& \left\{
    \begin{array}{ll}
    -\lambda\cos\theta_0\,\varepsilon_k, & \eta \ll 1, \\ 
    -\lambda\,i\omega, & \eta \gg 1.
    \end{array}
    \right.
\label{cases} 
\end{eqnarray}

When the system approaches QCP, and $\xi$ increases, the parameter
$\eta \propto \xi^2$ changes from $\eta\ll 1$ to $\eta\gg 1$.  The
crossover between the two asymptotic limits in Eq.\ (\ref{cases})
takes place at $\eta\sim1$, which corresponds to
\begin{equation}
\lambda\sim\lambda_{\rm cr} = \sqrt{g/E_F} \ll 1.
\label{lambdacr}
\end{equation}
Thus, the upper line in Eq.\ (\ref{cases}) stops being applicable
already at $\lambda\ll1$, before $\lambda$ can generate a divergence
in Eq.\ (\ref{pert}).  In the vicinity of QCP, the lower line in Eq.\
(\ref{cases}) applies, and, instead of Eq.\ (\ref{pert}), we find
\begin{equation}
\label{MASS}
  {m^* \over m} \approx {1 \over Z} \approx 1 + \lambda.
\end{equation}
We see therefore that the effective mass in Eq.\ (\ref{MASS}) diverges
only at QCP, where $\xi\to\infty$, but not before, contrary to the
conclusion one could draw from the Hartree-Fock approximation. This is
the central result of the paper.  Notice that the requirement $g>0$,
mentioned after Eq.\ (\ref{chi}), guarantees that $Z\leq1$, because
${\rm sgn}(\lambda)={\rm sgn}(g)$.

Further, since $\lambda_{\rm cr}\ll1$, vertex corrections and
renormalization of the fermionic $G$ in Eq.\ (\ref{S1}) are small at
$\lambda\sim\lambda_{\rm cr}$ and can be safely neglected.  This
justifies our approximation of including only the renormalization of
the bosonic propagator.  Moreover this approximation actually remains
valid even at larger $\lambda\agt1$.  Indeed, the modifications to
Eq.\ (\ref{S1}) due to vertex corrections and residual momentum
dependence of $\Sigma$ are small in the parameter $\sqrt{g/E_F}\ll1$
and can be safely neglected even when $\lambda = O(1)$. Although the
fermionic $\Sigma (\omega)$ is not small at $\lambda = O(1)$, using
the renormalized Green's function
$G^{-1}(k,\omega)=i\omega(1+\lambda)-v_F
(k-k_F)=Z[i\omega-v_F^*(k-k_F)]$ in Eq.\ (\ref{S1}) does not modify
the lower line in Eq.\ (\ref{cases}), because the extra factor $Z$ and
the renormalization of $v_F^*=k_F/m^*$ compensate each other
\cite{Kadanoff}.  Similarly, the coefficient $\gamma$ in Eq.\ 
(\ref{chi}) does not change, because the factor $Z^2$ coming from the
two Green's function in the polarization bubble (\ref{P}) compensates
the renormalization of the factor $1/v_F^2\to1/(v^*_F)^2$ in the
expression (\ref{chi}) for $\gamma$.  This behavior is typical for the
Migdal-Eliashberg-type theories \cite{Chubukov,Dirk}.

\paragraph{Anomaly in the calculation of self-energy.}  

Now we present details of the self-energy calculation and also explain
why $\lambda_{\rm cr}$ vanishes if the fermionic bandwidth ($\sim
E_F$) is set to infinity.  Linearizing the fermionic dispersion near
the two hot spots, we introduce $\zeta=v_F\tilde q_\perp$ and
$\tilde\epsilon_k=(\bm k-\bm k_F)\cdot\bm v_{\bm k_F+\bm
q_c}=\varepsilon_k\cos\theta_0$, where the vector $\bm k_F$ is
selected parallel to $\bm k$ (see Fig.\ \ref{fig:circle}).  Then
$\Sigma(k,\omega)\equiv\Sigma(k,\omega)-\Sigma (k_F,0)$ is
\begin{eqnarray}
  \Sigma(k,\omega)&=&-(i\omega-\tilde\epsilon_k)\,I(k,\omega),
\label{ne11} \\
  I(k,\omega) &=&\int
  \frac{d\Omega\,d\zeta\;\;{\tilde V}(\Omega, \zeta)}
  {2\pi\,[i(\omega+\Omega)-\tilde\epsilon_k-\zeta]\,(i\Omega-\zeta)},
\label{I}
\end{eqnarray}
where we introduced ${\tilde V}(\Omega, \zeta)$ similarly to Eq.\
(\ref{lambda})
\begin{equation}
  {\tilde V}(\Omega, \zeta)=\frac{2}{v_F}
  \int\frac{d\tilde q_{\|}}{(2\pi)^2}\, 
  V(\tilde q_\perp,\tilde q_\|,\Omega)
  =\frac{\lambda}{\sqrt{1+\gamma|\Omega|\xi^2}}.
\label{chi_eff}
\end{equation}
Notice that, to this accuracy, ${\tilde V}(\Omega,\zeta)$ does not
depend on $\zeta$, i.e., ${\tilde V}(\Omega,\zeta)={\tilde
V}(\Omega)$.

The evaluation of $I(k,\omega)$ in the limit $k\to k_F$ and
$\omega\to0$ requires care, because the integrand in Eq.\ (\ref{I})
contains two closely located poles separated by $\omega$ and
$\tilde\epsilon_k$.  If we approximate ${\tilde V}(\Omega)$ by a
constant ${\tilde V}(0)=\lambda$, then, nominally, the integral
(\ref{I}) is ultraviolet-divergent and depends on the order of
integration over $\Omega$ and $\zeta$.

To evaluate the integral correctly, one must keep in mind that Eq.\
(\ref{I}) is approximate, and higher-order terms in $(\bm q-\bm q_c)$
in G and $V$ always make the integral over $\bm q$ convergent at
$q-q_c\sim k_F$.  If $\gamma=0$ in Eq.\ (\ref{chi_eff}), then the
integral over $\Omega$ must be taken first, because its convergence is
provided only by the fermion Green's functions in Eq.\ (\ref{I}).  In
this case, we obtain $I(k,\omega)=\lambda\tilde\epsilon_k/
(i\omega-\tilde\epsilon_k)$ and
$\Sigma(k,\omega)=\Sigma_k=-\lambda\tilde\epsilon_k$, reproducing the
top line of Eq.\ (\ref{cases}).

On other hand, if $\gamma$ is large in Eq.\ (\ref{chi_eff}), then
${\tilde V} (\Omega)$ strongly depends on $\Omega$ and provides
convergence of the integral over $\Omega$.  In this case, it is
appropriate to integrate over $\zeta$ first, over the region where the
linearized expression (\ref{I}) is valid.  Taking the integral over
$\zeta$ first, we obtain $I(k,i\omega)=\lambda
i\omega/(i\omega-\tilde\epsilon_k)$ and $\Sigma(k,\omega)=-\lambda
i\omega$, reproducing the bottom line of Eq.\ (\ref{cases}).  Notice
that, although the frequency dependence of ${\tilde V}(\Omega)$ is
essential to determine the correct order of integrations, the strength
$\gamma$ of this dependence drops out from the final answer.  This
situation bears mathematical similarity to the chiral anomaly in
quantum field theory \cite{anomaly,Haslinger}.

The crossover between these two cases takes place when the
characteristic $\Omega$ in Eq.\ (\ref{chi_eff}) becomes of the order
of $\zeta\sim E_F$. Using the definition (\ref{eta}), we find that the
cases of weak and strong frequency dependence correspond to
$\eta\alt1$ and $\eta\agt1$, as in Eq.\ (\ref{cases}).

The fact that the crossover occurs at $\eta\sim1$, i.e., at
$\lambda_{\rm cr}\ll1$, is a consequence of $V({\bm q},\Omega)$ being
peaked on a \emph{circle} $|{\bm q}|=q_c$.  If $V({\bm q},\Omega)$
were peaked at a given \emph{vector} ${\bm q}_c$ in a crystal, then
$\tilde{V}$ would have the conventional, Ornstein-Zernike form
${\tilde V}(\Omega,{\tilde q}_{\perp})=\int d\tilde
q_\|/(\xi^{-2}+\tilde q_\perp^2+\tilde q_\|^2+\gamma|\Omega|) \propto
(\xi^{-2}+\tilde q_\perp^2+\gamma|\Omega|)^{-1/2}$. In this case, the
crossover takes place when all three terms become comparable:
$\xi^{-2}\sim\tilde q_\perp^2\sim\gamma|\Omega|$.  Since typical
$\Omega\sim v_F\tilde q_\perp$, the crossover in the vector case
occurs at $\gamma v_F\xi\sim1$, i.e., $\lambda_{\rm cr}\sim1$
\cite{Chubukov}.  However, $m^*$ does not diverge even when $\Sigma$
remains $\Sigma (k)$ up to $\lambda\sim1$, because the correction to
velocity $\partial_{\bm k}\Sigma=-\lambda\bm v_{\bm k+\bm q_c}$ is not
antiparallel to $\bm v_{\bm k}$ in the absence of nesting, so the
magnitude of the Fermi velocity does not vanish for any finite
$\lambda$ \cite{Chubukov}.

For completeness, it is instructive to see how the crossover from the
top to the bottom line in Eq.\ (\ref{cases}) happens if we always
integrate over $\Omega$ first in Eq.\ (\ref{I}).  Let us deform the
contour of integration over $\Omega$ to either upper or lower complex
half-plane.  For $-\tilde\epsilon_k<\zeta<0$, each half-plane contains
just one pole.  Wrapping the contour around the pole and integrating
over $\zeta$ within the specified limits, we obtain the $\Sigma_k$
contribution to $\Sigma$.  If $\tilde V$ does not depend on $\Omega$,
the calculation stops here.  However, when ${\tilde V}(\Omega)$
depends on $\Omega$ and is given by Eq.\ (\ref{chi_eff}), we also need
to consider a contribution from the branch cut in ${\tilde V}(\Omega)$
along the imaginary axis of $\Omega$ where $\Omega=i\nu+\delta$ and
$|\Omega|=i\nu\,{\rm sgn}\,\delta$. Evaluating the contribution from
the branch cut and combining it with the contribution from the pole,
we find $\Sigma = -\lambda\tilde\epsilon_k +
(i\omega-\tilde\epsilon_k) I_{\rm bc}$, where
\begin{eqnarray}
 I_{\rm bc} &=& (2/\pi)\int_0^{E_F} d\zeta \int_0^\infty
  d\nu\,{\rm Im}\tilde V(i\nu)/(\nu+\zeta)^2 
\nonumber \\
  &=& {2\over\pi} \int_0^\eta dz \int_0^\infty
  \frac{d\varpi}{(\varpi+z)^2}\,
  {\rm Im}\left(\frac{\lambda}{\sqrt{1-i\varpi}}\right).
\label{bc}
\end{eqnarray}
Here we introduced dimensionless variables $z=\zeta\gamma\xi^2$ and
$\varpi=\nu\gamma\xi^2$.  For $\eta\ll1$, Eq.\ (\ref{bc}) gives a
small $I_{\rm bc}\sim\lambda\eta\ln(1/\eta)$.  In the opposite limit
$\eta\gg1$, the integral over $z$ can be extended to infinity, and the
integral over $\varpi$ yields $I_{\rm bc}={\tilde V}(0)=\lambda$ via
the Kramers-Kronig relation.  Notice that this result does not depend
explicitly on the detailed form of the frequency dependence in $\tilde
V(\Omega)$ as long as the integral (\ref{bc}) quickly converges and
can be extended to infinity.  Substituting $ I_{\rm bc}$ into
$\Sigma$, we reproduce both lines in Eq.\ (\ref{cases}) for $\eta\ll1$
and $\eta\gg1$.

\paragraph{Conclusions.}  

In this paper, we studied the quantum critical behavior of an
isotropic system of fermions near a $T=0$ transition into a
density-wave state with a finite momentum $q_c$.  We demonstrated
that, upon approaching QCP, the fermionic self-energy crosses over
from $\Sigma(k,\omega)\approx\Sigma(k)$ to
$\Sigma(k,\omega)\approx\Sigma(\omega)$.  We showed that the crossover
occurs while the dimensionless coupling $\lambda$ (which diverges at
QCP) is still small. We found that the effective mass remains finite
and positive away from QCP, and diverges only at QCP as
$m^*\propto1/Z\propto1+\lambda$.  

Our results apply to both charge- and spin-density-wave instabilities.
In the latter case, spin-orbital interaction generally induces
anisotropy in the spin space, e.g., for easy axis, the interaction in
Eq.\ (\ref{Hint}) is mediated by $z$ component of spins.  This only
affects the numerical coefficient in $\Sigma$, proportional to the
number of fluctuating spin components, but does not change the
conclusions.

We thank D. Maslov for useful discussions.  The work was supported by
the NSF Grant DMR-0240238 (AVC); by US-ONR, LPS, and DARPA (VMG); and
by the NSF Grant DMR-0137726 (VMY).

\vspace{-1.5\baselineskip}


\begin{thebibliography}{00}
\bibitem{g<0} Ref.\ \cite{Yakovenko} considered the phenomenological
model (\ref{Hint}) and (\ref{chi_st}) with $g<0$ and
$q_c>k_F\sqrt{2}$.  In this case, frequency dependence of
$V(q,\Omega)$ cannot be obtained from the Landau damping as in Eq.\
(\ref{chi}).
\end{thebibliography}
\end{document}